\documentclass[letterpaper, 10 pt, conference]{ieeeconf}
\IEEEoverridecommandlockouts                              
\usepackage{graphics} 
\usepackage{amsmath} 
\usepackage[dvipsnames]{xcolor}
\usepackage{graphicx}
\usepackage{listings}
\usepackage{xcolor}
\usepackage{lipsum}
\usepackage{cuted}
\usepackage[utf8]{inputenc}
\usepackage{xcolor}
\usepackage{multirow}

\usepackage{mathtools}
\usepackage{amsfonts}

\usepackage{hyperref}
\usepackage{url}

\usepackage{pgfplots} 
\usepackage{dblfloatfix}
\usetikzlibrary{patterns}
\DeclareMathOperator*{\concat}{%
    \mathchoice%
        {\Big\Vert}%
        {\big\Vert}%
        {\Vert}%
        {\Vert}%
}
\usepackage{hyperref} 
\hypersetup{       
    colorlinks=true,            
    citecolor=blue,
    linkcolor=red
}

\definecolor{codegreen}{rgb}{0,0.6,0}
\definecolor{codegray}{rgb}{0.5,0.5,0.5}
\definecolor{codepurple}{rgb}{0.58,0,0.82}
\definecolor{backcolour}{rgb}{0.95,0.95,0.92}

\lstdefinestyle{mystyle}{ backgroundcolor=\color{backcolour},   
commentstyle=\color{codegreen},
keywordstyle=\color{magenta},
numberstyle=\tiny\color{codegray},
stringstyle=\color{codepurple},
basicstyle=\ttfamily\footnotesize,
breakatwhitespace=false,         
breaklines=true,                 
captionpos=b,                    
keepspaces=true,                 
numbers=left,                    
numbersep=5pt,                  
showspaces=false,                
showstringspaces=false,
showtabs=false,                  
tabsize=2
}

\title{\LARGE \bf
Multi-class Brain Tumor Segmentation using Graph Attention Network 
}

\makeatletter
\newcommand{\linebreakand}{%
  \end{@IEEEauthorhalign}
  \hfill\mbox{}\par
  \mbox{}\hfill\begin{@IEEEauthorhalign}
}
\makeatother

\author{
        Dhrumil Patel \\
	    Computer Science\\
	    Lakehead University\\ 
        pateld72@lakeheadu.ca
        \and
        Dhruv Patel \\
	    Computer Science\\
	    Lakehead University\\ 
        pateld181@lakeheadu.ca
        \and        
        Rudra Saxena \\
	    Computer Science\\
	    Lakehead University\\ 
        rsaxena1@lakeheadu.ca\\
        \and 
        Thangarajah Akilan \\
	    Software Engineering\\
	    Lakehead University\\ 
        takilan@lakeheadu.ca
}

\begin{document}
\maketitle
\thispagestyle{empty}
\pagestyle{empty}

\begin{abstract}
Brain tumor segmentation from magnetic resonance imaging (MRI) plays an important role in diagnostic radiology. To overcome the practical issues in manual approaches, there is a huge demand for building automatic tumor segmentation algorithms. This work introduces an efficient brain tumor summation model by exploiting the advancement in MRI and graph neural networks (GNNs). The model represents the volumetric MRI as a region adjacency graph (RAG) and learns to identify the type of tumors through a graph attention network (GAT) -- a variant of GNNs. The ablation analysis conducted on two benchmark datasets proves that the proposed model can produce competitive results compared to the leading-edge solutions. It achieves mean dice scores of 0.91, 0.86, 0.79, and mean Hausdorff distances in 95th percentile (HD95) of 5.91, 6.08, and 9.52 mm, respectively, for whole tumor, core tumor, and enhancing tumor segmentation on BraTS2021 validation dataset. On average, these performances are $>6\%$ and $>50\%$, compared to a GNN-based baseline model, respectively, on dice score and HD95 evaluation metrics.  


\end{abstract}

\section{Introduction}\label{sec:intro}
A brain tumor, a.k.a. intracranial tumor, is an abnormal growth of tissue that multiplies uncontrollably, which is found to be unchecked by the normal cell-cycle control mechanism\footnote{\href{https://www.aans.org/en/Patients/Neurosurgical-Conditions-and-Treatments/Brain-Tumors}{The American Association of Neurological Surgeons (AANS)}}. 
The  national brain tumor society (NBTS) estimates that there are 700,000 people in the United States who have brain tumors\footnote{\href{https://braintumor.org/brain-tumors/about-brain-tumors/brain-tumor-facts/}{The National Brain Tumor Society (NBTS)}}. 
Brain tumors greatly affect the quality of human life since the average five-year relative survival rate of patients who have been diagnosed with malignant brain tumors is $35.6\%$, according to NBTS.
Gliomas are the most prevalent forms of brain tumors.
Since they are highly invasive and develop aggressively infiltrating the central nervous system, early detection of gliomas is imperative for enhancing patient care and tailoring treatment strategies and saving lives. 
Modern diagnostic radiology for brain tumors mainly uses MRI-based neuroimaging technology. MRI plays a vital role in various stages of a patient even after the diagnosis of brain tumor, including surgical and radiotherapy treatment planning (RTP), longitudinal tumor monitoring, and treatment response evaluation~\cite{ostrom2019cbtrus, pandya2020multi, huo2015label}.
Since brain tumors can occur anywhere in the brain and can vary greatly in size, form, and morphology, 
the task of MRI brain tumor segmentation remains a grueling task even with expert's assistance. This becomes more challenging when the 
MRI scans are distorted or poorly contrasted due to non-uniformity in the static magnetic field and non-linearity in the gradient magnetic field~\cite{yasuda2016study}. To alleviate these challenges, researchers have shifted their focus to building data-driven intelligent assistive technologies (IATs) to automatically detect, and segment brain tumors from MRI scans. 

In this direction, this work proposes a GAT-based multi-class brain tumor segmentation model by exploiting multi-parametric MRI (mpMRI) that constitutes modalities, like native T1-weighted (T1), contrast-enhanced (T1CE), T2-weighted (T2), and T2-fluid-attenuated inversion recovery (FLAIR). 
GATs, in contrast to earlier approaches, enable simultaneous processing of the entire brain region and explicitly account for both local and global information by combining data from nearby nodes in the graph representation of the inputs. 
As a result, GATs have a rich mechanism for feature extraction and reasoning power to capture relationships among the data points efficiently. 
The main  objective of this work is to develop an algorithm that can perform accurate brain tumor segmentation from unstructured MRI modalities. It includes two key contributions: (i) a computationally efficient GAT model for brain tumor segmentation as compared to some of the recent state-of-the-art methods, and (ii) exhaustive experimental analysis on benchmark datasets and thorough comparative analysis.
The remainder of this paper is structured as follows.
Section~\ref{literature review} describes the related works, Section~\ref{proposed model} elaborates the proposed solution, Section~\ref{experimental anaylsis} discusses the exhaustive experiments and comparative analysis, and Section~\ref{conclusion} concludes the paper.

\section{Literature Review}\label{literature review}
\begin{figure*}[!hbp]
\centering
  \includegraphics[trim={0.5cm, 0.6cm, 0.4cm, 0.6cm},clip, width=1.0  \textwidth]{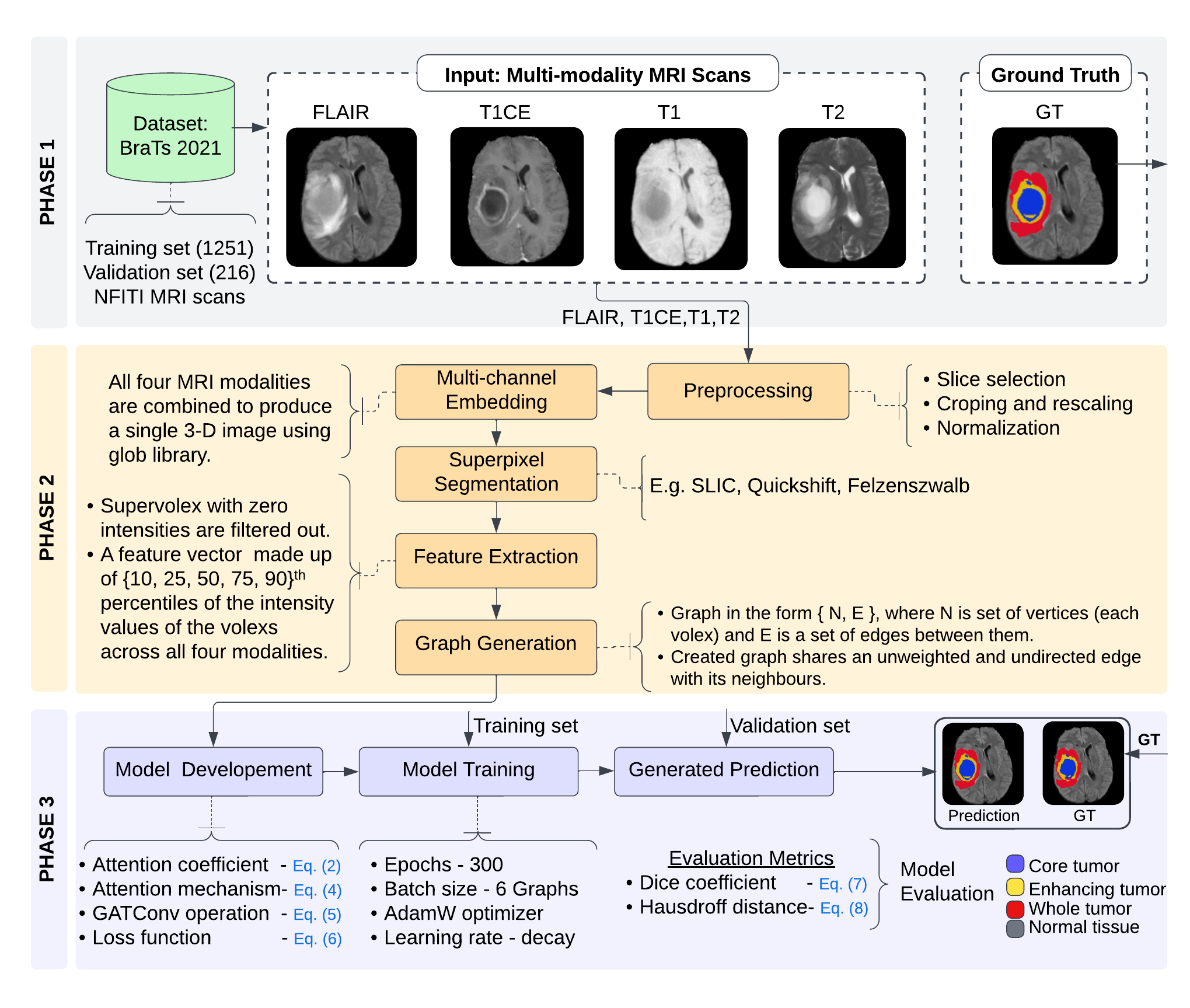} 
\caption{Detailed functional flow diagram of the proposed GAT-based MRI brain tumor segmentation framework. It consists of three abstract phases -- \textbf{Phase 1}: Data input, \textbf{Phase 2}: Data curation, and \textbf{Phase 3}: Model building and evaluation. For in-depth information, refer to Section~\ref{proposed model}.
} \label{fig:Detailed}
\end{figure*}

In the early 2000s, MRI brain tumor segmentation was predominately conducted using traditional machine learning (ML) measures to quantify tumors for clinical purposes. For instance, Shubhangi \textit{et~al.}~\cite{shubhangi2009support} proposed a solution with SVM combining knowledge-based approaches and multi-spectral analysis. 
On the other hand, some researchers attempted to integrate clustering and classification techniques for tumor detection and classification. Gopal\textit{et~al.}~\cite{gopal2010diagnose}, for example, proposed MRI brain tumor detection using fuzzy c-means (FCM) clustering with particle swarm optimization (PSO).
Similarly, Jahanavi and Kurup~\cite{7807881} introduced a hybrid model that integrates an SVM classifier with two clustering techniques k-means and  fuzzy c-mean. 
Javaid~\textit{et al.}~\cite{javaid2020hybrid} came up with an automated tumor recognition model using a kernel-based fuzzy c-means and skull stripping method. They used multiple kernels to extract spatial features for improved clustering and segmentation of tumors. 

Although there are several attempts in using traditional approaches, they all demonstrate one common limitation -- low reliability and poor performance in handling 3-D MRI data. To alleviate this, convolutional neural network (CNN)-based image segmentation models, like FCN~\cite{long2015fully}, encoder-decoder architectures~\cite{ronneberger2015u, badrinarayanan2017segnet, akilan2019video} have been exploited, in recent times. However, applying CNN on mpMRI is computationally burdensome.
This issue can be circumvented by representing the multi-modality MRI scans as graphs, which reduces the complexity from millions of voxels to several thousand nodes while preserving key information. Using such graph representation one can build efficient GNNs~\cite{gorinew} for various problems, like segmentation as in this work. 
Spectral-based GNNs have been successfully applied for image semantic image segmentation tasks~\cite{noh2015learning}. These GNNs capture long-range contextual features using two orthogonal graphs. The first graph learns spatial correlations between pixels, while the second graph simulates dependencies between feature maps of the network. Hyeonwoo~\textit{et al.}~\cite{noh2015learning} combined these multi-view representations using multiple simultaneous self-constructing graph modules (SCG) and graph convolutional neural networks (GCNs). 
Saueressig~\textit{et al.}~\cite{saueressig2020exploring} and Wei~\textit{et~al.}~\cite{wei2022predicting} investigated a GNN model for automatic brain tumor segmentation. 
Wei~\textit{et~al.}~\cite{wei2022predicting} proposed a method to determine isocitrate dehydrogenase mutation status in glioma using structural brain networks and GNNs. These GNN-based models have shown competitive performances when compared to 3d-CNN and 3d-DenseNet-based segmentation approaches. 
To improve the performance of GNN-based brain tumor segmentation, this work exploits the attention mechanism of GATs. In GNNs, GAT has been one of the most powerful cutting-edge architectures~\cite{velivckovic2017graph}.



\begin{figure*}[!htp]
\centering
  \includegraphics[trim={0.2cm 0.30cm 0.2cm 0.2cm},clip, width=0.9\textwidth]{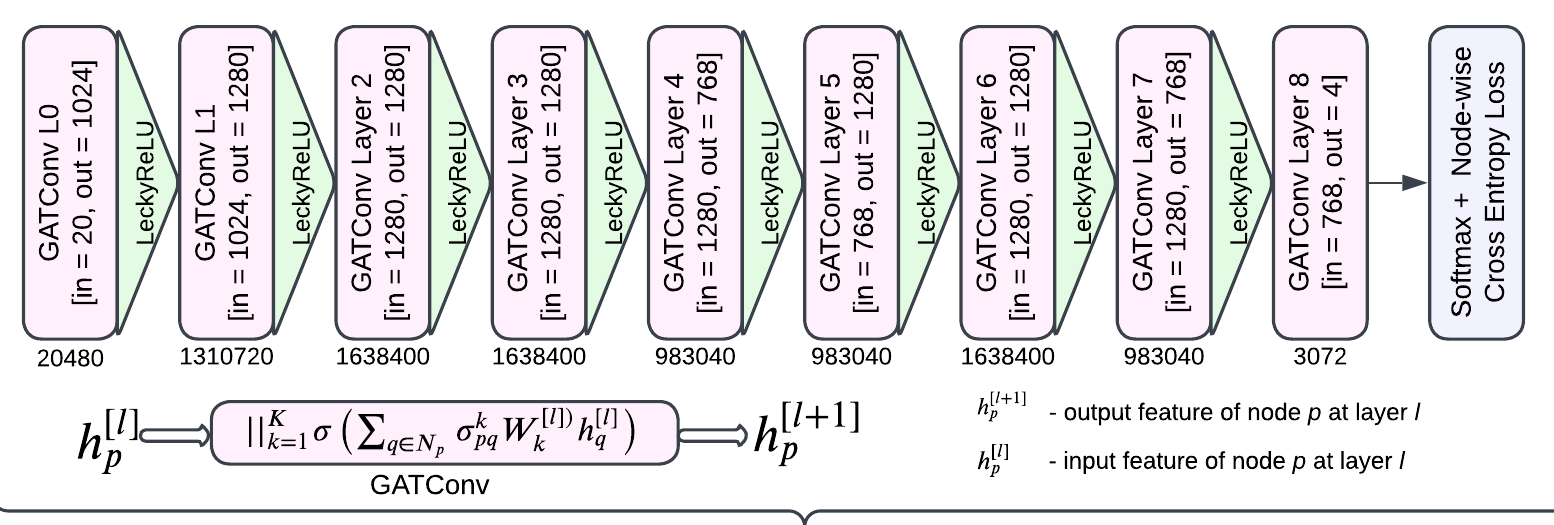}
  \caption{Layer-wise schematic of the proposed GAT containing 10,208,524 total number of trainable parameters. It stacks eight GATConv layers with LeakyReLu activation, and the top layer is formed by a GATConv layer with Softmax activation. Each GATConv layer computes each node's output features as shown in GATConv module using \eqref{eqn:eq7}, where $h_{u}^{[l]}$ - input features, and  $h_{u}^{[l+1]}$ - output features. LeakyReLU uses a negative input slope, $\eta = 0.2$ while applying the attention mechanism defined in \eqref{eqn:eq6}. The model is trained to minimize the node-wise multi-label cross-entropy loss given by \eqref{eqn:eq8}. The values denoted below each layer are the number of trainable parameters of the respective layer. } 
  \vspace{-0.2cm}
\label{fig:layerdiagram}
\end{figure*}

\section{Proposed GAT-based Solution}\label{proposed model}

Fig.~\ref{fig:Detailed} on page~\pageref{fig:Detailed} depicts the process flow of the proposed multi-class brain tumor segmentation framework that subsumes three phases: input, data curation, and model building.

\subsection{Phase 1: Input}\label{A_phase1}

The BraTS 2021~\cite{menze2014multimodal} mpMRI scan dataset is used as the input source. It provides four MRI sequences, including T1, T1CE, T2, and FLAIR, along with ground truth annotations for three tumor types -- core tumor, enhancing tumor, and whole tumor~\cite{pandya2020multi, menze2014multimodal, bakas2018identifying, baid2021rsna}. 

\subsection{Phase 2: Data Curation}
\subsubsection{Data preprocessing}

All mpMRI scans are standardized through the conversion of DICOM images into neuroimaging informatics technology initiative (NIfTI) format, co-registering to the same anatomical template (SRI24), re-sampling to a uniform isotropic resolution (1mm3), and skull-stripping~\cite{baid2021rsna}. 
Hence, the mpMRI scans are padded to a uniform shape and region cropped  to a precise bounding box around the brain. 
Then, each scan's intensities are re-scaled by dividing them by their 99.5th percentile raw intensity value, so the intensity range becomes [0, 1]. 
Finally, each modality is refined to have a zero mean and unit variance. 

\subsubsection{Superpixel segmentation}

All four MRI modalities are combined to produce a single 3-D image, from which the RAG is generated. Here, simple linear iterative clustering (SLIC) algorithm~\cite{achanta2012slic} is applied to the 3-D image, to get a set of $k$ supervoxels (a volume of superpixel segmentation).
To group supervoxels with similar intensity values and spatial locations, SLIC employs K-means with Euclidean norm, $D_{s}$ between two voxels $p$ and $q$ as defined in \eqref{eqn:eq1}. 
This operation reduces the number of supervoxels considerably. 
\begin{equation}
\begin{split}
     \resizebox{.91\hsize}{!}{$D_{i}=\sqrt{\left(\alpha_{p}-\alpha_{q}\right)^2 + \left(\beta_{p}-\beta_{q}\right)^2 + \left(\gamma_{p}-\gamma_{q}\right)^2 + \left(\delta_{p}-\delta_{q}\right)^2}$} \\
  D_{xyz}= \sqrt{\left(x_{p} - x_{q}\right)^2 + \left(y_{p} - y_{q}\right)^2 + \left(z_{p} - z_{q}\right)^2},\\
    D_{s} = \sqrt{D_{i}^2 + \left(\frac{\omega}{\lambda}\right)^2D_{xyz}^2},
\end{split}\label{eqn:eq1}
\end{equation}
where $\alpha, \beta, \gamma, \delta$, and $\{x, y, z\}$ represent intensity values of the mpMRI scans T1, T1CE, T2, FLAIR, and a voxel's spatial position in image coordinate, respectively. Hence, $\lambda$ and $\omega$ are the parameters that control the supervoxel compactness and the anticipated space between them, respectively, based on intensity and spatial information~\cite{saueressig2020exploring, achanta2012slic}. 
This work follows two of the most recent state-of-the-art studies \cite{saueressig2020exploring} and \cite{saueressig2022joint} to set the  hyperparameters of SLIC optimally. Note that SLIC is found to be robust among many other existing methods to generate superpixels from image intensity maps.


\subsubsection{Feature extraction}\label{sec:fe}

In this stage, the outliers, i.e., the supervoxels outside of the brain mass, or having zero intensity are removed. 
Then, a feature vector size of twenty is created using the 10th, 25th, 50th, 75th, and 90th percentiles of the intensity values of each cluster across all four MRI modalities. 
Each of these 20-dimensional feature vectors is associated with a ground truth segmentation label. 


\subsubsection{Graph generation}


The RAG is generated by considering each superpixel cluster as a node and connecting edges between all adjacent clusters. The generated RAG is fed to the proposed GNN, more specifically, the GAT (cf.~Fig.~\ref{fig:layerdiagram}), where each node in the RAG is represented by its 20-dimensional feature vector and a label as described earlier. Note that during testing the feature vector does not include the label information.

\subsection{Phase 3: Model Building}\label{phase_3}
\subsubsection{The architecture}

Fig.~\ref{fig:layerdiagram} depicts a layer-wise schematic of the proposed GAT. The model transforms the input features through a linear operation followed by a nonlinear activation, and an attention function. The attention function gives a higher weightage to the edges between neighboring nodes that have the same label~\cite{velivckovic2017graph, thekumparampil2018attention}. 
To stabilize the learning process of self-attention, this work employs multi-head attention similar to the one used by Vaswani~\textit{et al.}~\cite{vaswani2017attention}. The computations carried out in each GATCov layer are summarized as follows (vivid descriptions can be found in \cite{velivckovic2017graph}). 
The GATCov layer takes a vector of node features, $\mathbf{h}^{[l]} = \{ \vec{h_{1}},\vec{h_{2}},\cdots, \vec{h_{n}}\}$, $\vec{h_{i}} \in \mathbb{R}^{f}$ and generates an output vector, $\mathbf{h}^{[l+1]} \in \mathbb{R}^{f^\prime}$, where $n$, and $f$ stand for the number of nodes, and the node feature dimension, respectively. 
The linear transformation is parametrized by a weight matrix, $W\in R^{F^\prime \times F}$. It is then used to estimate the node attention value, $e_{pq}$ using a shared attention mechanism $a$ as defined in \eqref{eq:attn}. It indicates the importance of node $q$ to node $p$, where $q$ is in the neighborhood ($N$) of node $p$, $q \in N_p$ in the RAG.
\begin{equation}\label{eq:attn}
    e_{pq} = a\left(W\vec{h_{p}} , W\vec{h_{q}}\right). 
\end{equation}
The nonlinear activation is a LeakyReLu as defined in \eqref{eq:LReLU}. 
\begin{equation}\label{eq:LReLU}
    LeakyReLu(x) = \left\{
    \begin{array}{rcl} 
    x & \mbox{for} & x \geq 0 \\ 
    \eta \times x & \mbox{for} & x < 0
    \end{array},
    \right\}
\end{equation}
where $\eta$ is the negative slope. 
The estimated attention value, $e_{pq}$ is normalized via a softmax activation given by \eqref{eqn:eq6} to make it comparable across all the nodes. 
\begin{equation}\label{eqn:eq6}
\resizebox{.9\hsize}{!}{    
    $\sigma _{pq} = \frac{exp\left(LeakyReLU\left(\vec{a^ {T}}\left[W\vec{h_{p}}\concat W\vec{h_{q}}\right]\right)\right)}{\sum _{k\in N_{p}}exp\left(LeakyReLU\left(\vec{a^{T}}\left[W\vec{h_{p}}\concat W\vec{h_{k}}\right]\right)\right)}$},
\end{equation}
where $a^{T}$ represents the transposition of the weight vector $\vec{a} \in R^{2F^\prime}$ and $\left |  \right | $ is the concatenation operation. 
After computing the normalized attention values from  $K$ independent attention mechanisms, their features are amalgamated to generate output feature representation as defined in \eqref{eqn:eq7}. 
\begin{equation}
    \label{eqn:eq7}
    \mathbf{h}_{p}^{[l+1]} =   {\concat^{K}_{k=1}}\sigma \left(\sum_{q\in N_p}^{}\sigma_{pq}^{k}W_{k}^{[l])}h_{q}^{[l]}\right),
\end{equation}
where $\mathbf{h}_{p}^{[l+1]}$ is the out features of node $p$ at layer $l$, $\left |  \right |$ is the concatenation of features computed by $K$ multi-head attention mechanisms each computing their own pairwise self-attention $\sigma_{pq}^{k}$ between each pair of neighboring nodes $p$ and $q$ based on ~\eqref{eqn:eq6}, and  $W_k^{[l]}$ is the linear transformation's weight matrix, $\sigma$  is LeakyReLU defeined in \eqref{eq:LReLU}.

\begin{table*}[!t]
  \centering
  \caption{Performance Of Various Models on BRATS2021 Validation Dataset and Their $\%$ of Improvement in Dice Coefficient (higher is better) And HD95 (lower is better) Compared to the Baseline Model Introduced in \cite{saueressig2022joint}. \\ Note: $\downarrow$ and $\uparrow$ Stand for `+' and `-' Improvement, Respectively. The Best Performances are Inked in \textcolor{blue}{Blue}}  
  \setlength{\tabcolsep}{9pt}
  \renewcommand{\arraystretch}{1.5}  
  \begin{tabular}{|l|c|c|c|c|c|c|c|c|c|c|}
    \hline 
    \multirow{2}{*}{\textbf{Models}} &
      \multicolumn{4}{c|}{\textbf{Dice Score (\%)}} &
      \multicolumn{4}{c|}{\textbf{HD95}} &
      \multicolumn{2}{c|}{\textbf{Avg. \% of Improvement}} \\\cline{2-11}
     & WT & TC & ET & Average & WT & TC & ET & Average & Dice Score & HD95\\\hline\hline
    GNN~\cite{saueressig2022joint} & 0.87 & 0.78 & 0.74 & 0.80 & 6.92 & 16.67 & 20.40 & 14.66 & \multicolumn{2}{c|}{\textbf{Baseline}}  \\
    \hline
    GNN-CNN~\cite{saueressig2022joint} & 0.89 & 0.81 & 0.73 & 0.81 & 6.79 & 12.62 & 28.20 & 15.87 & 1.3 $\uparrow$ & 8.3 $\downarrow$\\
    \hline
    3D CMM-Net~\cite{10.1007/978-3-031-08999-2_28} & 0.84 & 0.81 & 0.75 & 0.80 & 10.16 & 24.64 & 35.00 & 23.26 & 0 & 58.7 $\downarrow$\\
    \hline
    3D-UNet~\cite{inbook} & 0.87 & 0.76 & 0.73 & 0.79 & 6.29 & 14.70 & 30.50 & 17.16 & 1.3 $\downarrow$ & 17.1 $\downarrow$ \\
    \hline
    3D ResUNet~\cite{10.1007/978-3-031-08999-2_26} & 0.90 & 0.85 & 0.82 & 0.86 & 4.3 & 9.89 & 17.89 & 10.69 & 7.5 $\uparrow$ & 27.1 $\uparrow$ \\
    \hline
    DNN~\cite{10.1007/978-3-031-08999-2_27} & 0.90 & 0.84 & 0.81 & 0.85 & 7.3 & 22.32 & 19.58 & 16.4 & 6.3 $\uparrow$ & 11.9 $\downarrow$\\
    \hline
    GAT (this work) & 0.91 & 0.86 & 0.79 & 0.85 & 5.91 & 6.08 & 9.52 & 7.17 & \textcolor{blue}{6.3} $\uparrow$ & \textcolor{blue}{51.1 $\uparrow$}\\
    \hline
    \end{tabular}
    \label{tab:performance}
\end{table*}



\subsubsection{Model training}

\begin{figure}[!t]
\includegraphics [width=8.5cm]{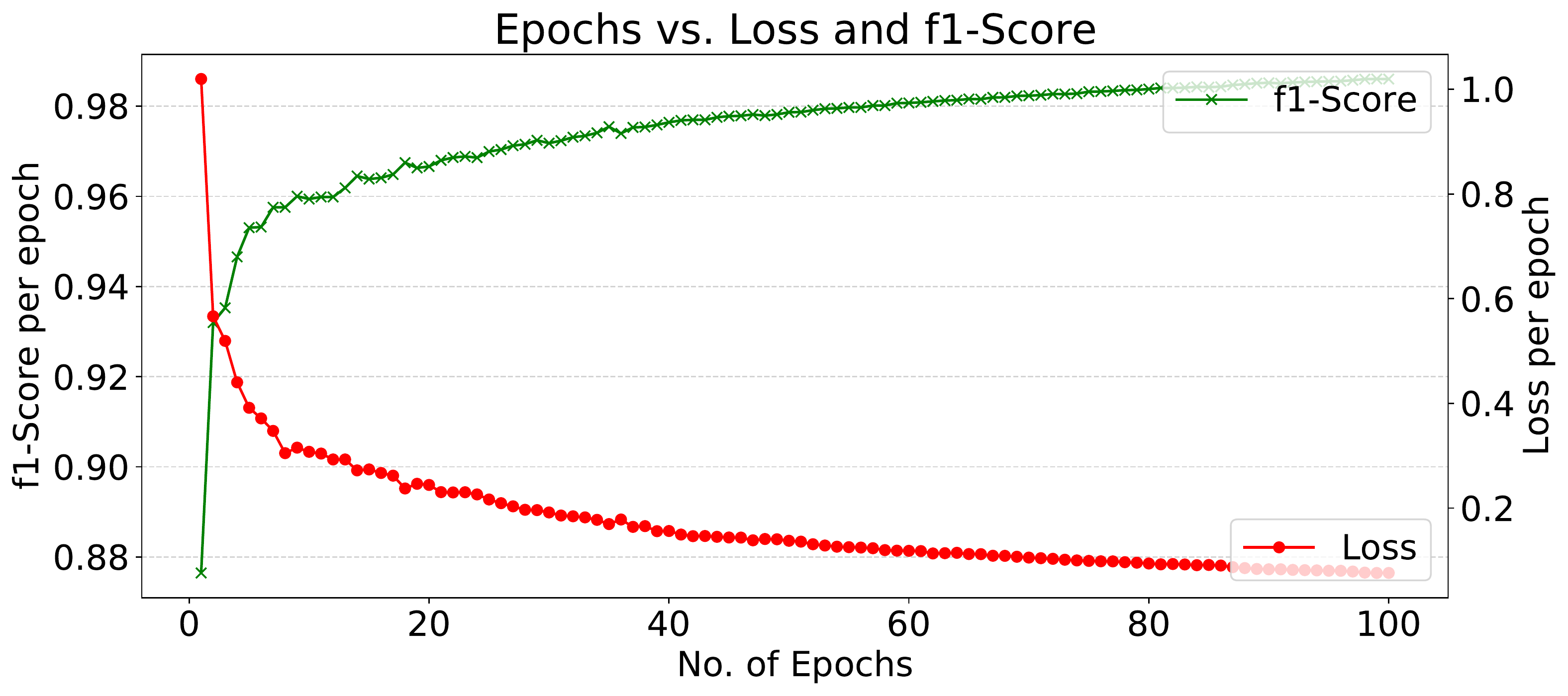} \vspace{-0.2cm}
\caption{Training progress of the proposed model in the first 100 epochs. \textcolor{Green}{Green line} indicates progression in \textcolor{Green}{f1-score} for whole tumor and the \textcolor{red}{Red line} indicates decrement in \textcolor{red}{node-wise multi-label cross-entropy loss}.
}\label{fig5}\vspace{-0.2cm}
\end{figure}

The model is trained using the Adam optimizer with exponential decay at the rate of 0.0001 to reduce the node-wise multi-label cross-entropy loss as in \eqref{eqn:eq8}. 
\begin{equation}
\label{eqn:eq8}
    Loss = \sum_{c=0}^{C}(1_{c=y})W_{c}\log(\widehat{P}_{y}),
\end{equation}
where $C$ are the possible classes, $W_{c}$ is the class weight, $y$ is the true label, $1_{c=y}$ is an indicator function, and $\widehat{P}_{y}$ is the predicted probability of the class label.
The model is trained for 300 epochs with mini-batches of six graphs taking approximately two days on the computational platform described in Section~\ref{sec:environment}. The training progress of the model for the first 100 epochs is shown in Fig.~\ref{fig5}.




\section{Experimental Analysis}\label{experimental anaylsis}
\subsection{Environment}\label{sec:environment}
The proposed solution is developed using Python 3 and its open-source native libraries along with deep learning libraries, PyTorch, and deep graph library (DGL). The model development, training, and testing are carried out on a system with an AMD Ryzen 7 4800HS 2.90 GHz processor connected to Google Colab with a Tesla K80 GPU having 2496 CUDA cores and 35 GB of DDR5 VRAM. 

\subsection{Evaluation Metrics}
Dice score~\eqref{eq:dice} and HD95~\eqref{eq:hd95} are used to assess the performance of the proposed model. 
While the dice score evaluates the degree to which predicted segmentation and the actual segmentation overlap, HD95 measures the degree to which they differ from each other.
\begin{equation}\label{eq:dice}
    Dice = \frac{2TP}{2TP + FP + FN},
\end{equation}
where $TP$, $FP$, and $FN$ are the number of true positives, false positives, and false negatives, respectively. 
\begin{equation}\label{eq:hd95}
    HD95 = 95\%\left(d\left(\widehat{Y},Y\right)\concat d\left(Y,\widehat{Y}\right)\right),
\end{equation}
where $d$ is the element-wise distance of every voxel in $\widehat{Y}$ to the closest voxel of the same label in $Y$. Here, $\widehat{Y}$ is the predicted label of each voxel, $Y$ is the ground truth label of each voxel, and $\left |  \right |$ is the concatenation operator.

\subsection{Performance Analysis}
\subsubsection{Quantitative analysis}
Table~\ref{tab:performance} on page~\pageref{tab:performance} provides a comprehensive comparative analysis of recent existing solutions with ours, where a GNN-based model introduced in \cite{saueressig2022joint} is considered as the baseline. 
The proposed model provides competitive results compared to the existing methods and achieves an average of $6\%$ and $51\%$ improvements over the baseline model in dice score and HD95. 
On average, it outperforms all the existing solutions wrt HD95 by achieving 7.16 mm less distance than the baseline. These results indicate that the addition of the multi-head attention mechanism can segment multi-class brain tumors more precisely.  
Hence, the average per-sample prediction time of the model is 1.7 seconds in the environment described earlier.

\subsubsection{Qualitative analysis}

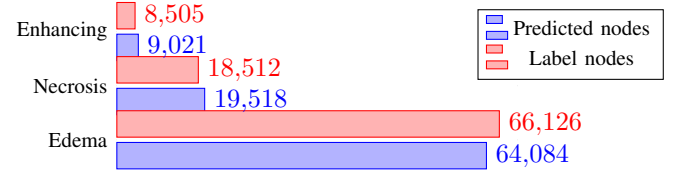
\begin{figure}[!tp]
\pgfplotsset{width=.8\columnwidth,
    height=0.5\columnwidth, every tick label/.append style={font=\footnotesize}}
\label{count}
\begin{tikzpicture}
    \begin{axis}
    [
    legend style={at={(0.9,0.7)},
          anchor=west,legend columns=1},
          legend style={font=\footnotesize},
    xbar,
    y axis line style = {opacity = 0},
    axis x line = none,
    tickwidth = 0.05pt,
    ytick distance=1,       
    label style={font=\footnotesize},
    tick label style={font=\footnotesize},
    enlarge y limits = 0.45,
    enlarge x limits  = 0.05,
    nodes near coords,
    symbolic y coords = {Edema, Necrosis, Enhancing},    
    ]
  \addplot coordinates { (64084,Edema) (19518,Necrosis) (9021,Enhancing) };
  \addplot coordinates { (66126,Edema) (18512,Necrosis) (8505,Enhancing) };
  \legend{Predicted nodes, Label nodes};
  \end{axis}
\end{tikzpicture} \vspace{-0.8cm}
\caption{Comparison of the actual label node counts and predicted node counts on 
 the BraTS2021 validation dataset the three types of tumors.}
\label{fig:my_label}
\end{figure}

\begin{figure*}[!tbp]
\rotatebox[origin=c]{-90}
  {\includegraphics [trim={2cm, 1cm, 0cm, 16cm},width=9.2cm]{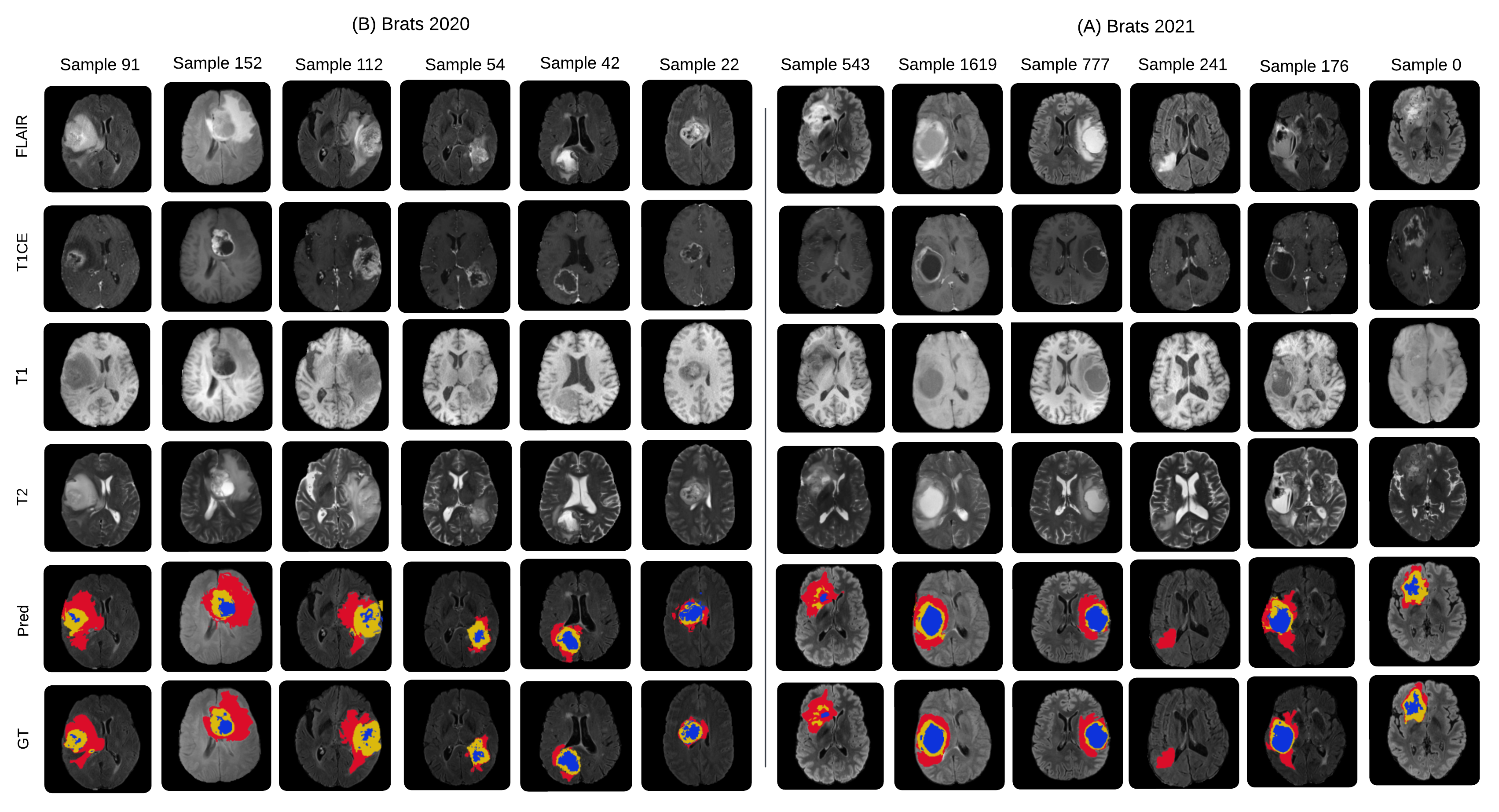}}
  \vspace{-0.5cm}
 \caption{Sample qualitative results of the proposed model. Rows 1 - 4 show the four input MRI modalities: FLAIR, T1CE, T1, and T2. Row 5 and 6 show the predicted tumor segmentation and its ground truth. Columns 1 - 6 show the used test samples from the BraTS2020 dataset, while columns 7 - 12 show the samples from BraTS2021  represent the segmentations obtained using the BraTS2020 validation dataset. Legend: \textcolor{red}{Red} - Edema (whole tumor), \textcolor{yellow}{Yellow} - Enhancing tumor, \textcolor{blue}{Blue} - Necrosis (core). Note: Here the BraTS2020 dataset is used under extended experiments to check the robustness of our model.
} \label{fig:viualization}
\end{figure*}

While Fig.~\ref{fig:my_label} on page \pageref{fig:my_label} compares the actual label node counts and predicted node counts on the BraTS2021 validation dataset for the three types of tumors: enhancing, necrosis, and edema, Fig.~\ref{fig:viualization} on page \pageref{fig:viualization} summarizes a few qualitative results of the proposed model on BraTS2021 validation dataset. We also conduct an extended experiment using the BraTS2020 dataset to check the robustness of our model. Note that the model has not been retrained on BraTS2020, rather the trained model on BraTS2021 is used to predict the tumor segmentation on BraTS2020. These qualitative results prove that the proposed GAT segments the three types of tumors close to the ground truth segments. 


\section{Conclusion}\label{conclusion}

Exploiting advancement in AI and computer vision for lifesaving medical diagnosis is extremely significant.
In response to that, this work proposes a graph attention-based neural network for effectively segmenting multi-class tumors from multi-modality MRI scans. 
The exhaustive experimental studies and comparative analysis on the benchmark BraTS2021 dataset shows that the proposed model can achieve competitive performances. It shows an overall improvement $>6\%$ and $>50\%$, respectively for dice score and HD95 evaluation metrics compared to an existing GNN-based baseline model.  

\bibliographystyle{ieeetr}
\bibliography{references}

\end{document}